\documentclass[aps,
		prd,
		reprint,
		twocolumn,
		superscriptaddress,
		shortbibliography,
		nofootinbib,
		floatfix,
		notitlepage
		]{revtex4-1}
\pdfoutput=1
\usepackage{amsmath,amssymb,amsfonts}

\usepackage{graphicx}
\usepackage{bbold}
\usepackage{slashed}
\usepackage{color}
\usepackage[colorlinks=true,citecolor=blue]{hyperref}
\hypersetup{colorlinks=true,citecolor=red,linkcolor=blue}

\newcommand{\be}{\begin{equation}}
\newcommand{\ee}{\end{equation}}
\newcommand{\bi}{\begin{itemize}}
\newcommand{\ei}{\end{itemize}}
\newcommand{\bea}{\begin{eqnarray}}
\newcommand{\eea}{\end{eqnarray}}

\newcommand{\ud}{\mathrm{d}}
\newcommand{\LCm}{{\scriptscriptstyle -}} 
\newcommand{\LCp}{{\scriptscriptstyle +}}
\newcommand{\LCpm}{{\scriptscriptstyle \pm}}

\newcommand{\LCperp}{{\scriptscriptstyle \perp}}

\newcommand{\bracket}[2]{\bra{#1}\,#2\rangle} 
\newcommand{\bra}[1]{\langle\,#1\,|}          
\newcommand{\ket}[1]{|\,#1\,\rangle}          

\usepackage[T1]{fontenc} 

\begin{document}
\title{The physics of adiabatic particle number in the Schwinger effect}
\author{Anton~Ilderton}
\email{anton.ilderton@ed.ac.uk}
\affiliation{Higgs Centre, School of Physics \& Astronomy, University of Edinburgh, EH9 3JZ, UK}
\begin{abstract}
The production of electron-positron pairs from light is a famous prediction of quantum electrodynamics.  Yet it is often emphasised that the number of produced pairs has no physical meaning until the driving electromagnetic fields are switched off, as otherwise its definition is basis-dependent. The common adiabatic definition, in particular, can predict the `creation' of a number of pairs orders of magnitude larger than the final yield. We show here, by clarifying exactly what is being counted, that the adiabatic number of pairs has an unambiguous and physical interpretation. As a result, and perhaps contrary to expectation, the large numbers of pairs seen at non-asymptotic times become, in principle, physically accessible.
\end{abstract}
%
%

\maketitle
\section{Introduction}
One of the most well-known \textit{non}-perturbative predictions of quantum electrodynamics is the Schwinger effect, the creation of electron-positron pairs from light~\cite{Heisenberg:1936nmg,Schwinger:1951nm}. An example of quantum tunnelling (from the Dirac sea), the Schwinger effect can be the dominant mechanism behind charge loss from black holes~\cite{Gibbons:1975kk}, exhibits features of universality~\cite{Gies:2015hia,Gies:2016coz},
and has analogues in solid-state physics~\cite{Schutzhold:2009fic,Klar:2019yxw}, often realised through Landau-Zener tunneling~\cite{Fillion-Gourdeau:2015dga,Linder:2015fba,Pineiro:2019uzb,Solinas:2020woq}. Experimental progress is bringing us closer to the point at which the Schwinger effect may be seen in laser experiments~\cite{Bulanov:2010ei,PhysRevA.85.033408,FedotovLimit,Gonoskov:2013ada,Otto:2016xpn,Torgrimsson:2016ant}.

There exist many, mutually consistent methods by which to calculate the asymptotically late time (final-state) number of pairs which can be produced from a given electric field profile~\cite{Holstein:1999ta,Dunne:2005sx,Dunne:2006st,Hebenstreit:2009km,Dumlu:2009rr,Hebenstreit:2010vz,Gelis:2015kya,Gavrilov:2016tuq,Torgrimsson:2018xdf}. However, attempts to define a time-dependent number of created pairs, $N(t)$, while the electric field is still turned on, runs into trouble: $N(t)$ depends unphysically on the choice of basis, and while all choices agree on the asymptotic number of pairs, they differ greatly at intermediate times. Even the most common
and well-known choice, that of a basis of adiabatic Hamiltonian eigenstates, yields pair numbers which fluctuate wildly in time, and can exhibit transient values orders of magnitude higher than the final, unambiguous, number of pairs. This has led to drastic overestimates for the number of pairs which could be created in experiments, and as such it is now repeatedly emphasised that no direct physical meaning should be attributed to the non-asymptotic number of pairs~\cite{Hebenstreit:2008ae,Kim:2011jw,Blinne:2013via,Zahn:2015awa}. (The issue is exemplified by `solitonic' cases for which $N(t)$ is non-zero, but falls to \textit{exactly} zero asymptotically, meaning no pairs are ultimately produced~\cite{Kim:2011jw,Huet:2014mta,Kim:2011sf}.)

However, it is also possible to find particular `super-adiabatic' bases for which $N(t)$ interpolates more smoothly between 0 and its asymptotic value, without the large oscillations of other bases~\cite{Dabrowski:2014ica,Dabrowski:2016tsx}. What, then, is the `correct' basis to use in order to describe the time evolution of the number of pairs~\cite{Dabrowski:2016tsx}? Do the transient excitations have a physical meaning~\cite{Blaschke:2013ip}, or should a method be found to remove them~\cite{Zahn:2015awa}?
{The same questions apply to analogue Schwinger effects, where e.g.~adiabatic particle number is also used, and related questions arise in particle creation from spacetime curvature~\cite{Parker:1969au,Fulling:1979ac,Parker:2012at,Yamada:2021kqw}, in chaos~\cite{Cooper:1994zzb,Faccioli:1997qa,Cooper:1998ae} and in tunnelling ionisation~\cite{Buttiker,Landsman,Camus-Wigner,Zimmermann}.}
Progress in understanding intermediate particle number would thus shed light on a non-perturbative quantum phenomenon of relevance to laser, condensed-matter, gravitational and nuclear physics.


{Our aim in this paper is to motivate a change in perspective:  rather than asking which of the infinitely many bases is physically relevant for pair production (if any), we will here turn the question around, and ask instead \emph{what is the physics contained in different bases}? We suggest that this may be the more physically relevant, and revealing, question. We will here identify the physical meaning of adiabatic particle number, by clarifying exactly what it counts.}


This paper is organised as follows. In Sec.~\ref{sec:setup} we describe our physical setup and conventions. {We will work throughout with time-dependent but spatially homogeneous fields; while these are not directly relevant to future (e.g. laser) experiments, they are the prototype example used in the study of non-perturbative pair production, beyond the completely constant case.} In Sec.~\ref{sec:adiabatic} we introduce the adiabatic basis and identify its physical interpretation. {We also provide some context for our results by applying them to the problem of pulse shaping to increase pair yields (though our aim is not to identify an optimum experimental setup or to go too far into phenomenology).} We conclude in Sec.~\ref{sec:outro}.

\section{Pair production from electric fields}\label{sec:setup}
%
We consider electron-positron pair production from an electric field $E(t)$ in $1+1$ dimensions for clarity. {This is the simplest setup with the correct (fermionic) statistics.} Working in the Schr\"odinger picture will yield simple expressions. We set $\hbar=c=1$, while $m$ and $e$ are the electron mass and charge.  $E(t)$ can be represented by the potential $A_0=0$ and $\partial_t A_1(t) = E(t)$, with $A_1(-\infty)=0$. {We write $a(t) := eA_1(t)$. Starting from the usual Hamiltonian density $\bar\psi(x) \big[m - i\gamma^1 (\partial_1 +ia(t))\big]\psi(x)$ for the fermion $\psi$, the Hamiltonian may be written}
\be
\nonumber
	H(t) \!= \!\int\!\!\ud p\, \Omega_\LCp(t) \big(b^\dagger_p b_p + d^\dagger_{-p}d_{-p} \big) + \Omega_\LCm(t) (b^\dagger_p d^\dagger_{-p} + d_{-p}b_p) \;,
\ee
in which the $\gamma$-matrices have been represented in terms of Pauli matrices as $\gamma^0 = \sigma^1$, $\gamma^1 = i\sigma^2$, the mode operators obey $\{b_p,b^\dagger_q\} = \{d_p,d^\dagger_q\} = \delta(p-q)$ and, in terms of the classical electron momentum $\pi(t) := p - a(t)$ in the background, we have {$\Omega_\LCp(t) = (\pi(t) p +m^2)/p_0$ and $\Omega_\LCm(t) = -ma(t)/p_0$.}  Energies are $\sqrt{p^2+m^2} = p_0$ and $\sqrt{\pi^2+m^2} = \pi_0$ as usual. 
%
%
%
%
%

%

\subsection{The vacuum before and after}
%
We begin in the vacuum $\ket{0}$, and turn on the electric field $E(t)$ at $t=t_0$, which may be finite or $-\infty$. The time-evolved state $\ket{0;t}$ obeying the Schr\"odinger equation $i\partial_t \ket{0;t}= H\ket{0;t}$  is
\be\label{sol1}
	\ket{0;t} := \exp\bigg[ - i V \vartheta(t) + \int\!\ud p \; \Omega_p(t) b^\dagger_p d^\dagger_{-p}\bigg] \ket{0} \;,
\ee
in which $V$ is the spatial volume (over $2\pi$), $\vartheta(t)$ is most easily determined by overall normalisation, while the `covariance' $\Omega(t)$ obeys $\Omega_p(t_0) = 0$ and
\be\label{schro}
\begin{split}
	i  {\dot \Omega_p(t)} &= 2 \Omega_\LCp(t) \Omega_p(t) + \Omega_\LCm(t)\big(1-\Omega_p^2(t)\big) \;.
\end{split}
\ee
%
{After the electric field turns off, the potential $a(t)$ either goes to zero or a nonzero constant $a_\infty$ (we will see examples of both cases)}. This is the same as a pure gauge background. {It is crucial for what follows to understand the physics in the corresponding} vacuum state $\ket{\mathbb{0}}$ obeying $H\ket{\mathbb{0}} = 0$ (there  is only free physics in a pure gauge background). From (\ref{schro}) the covariance in the pure gauge vacuum then obeys $0 = 2 \Omega_\LCp \Omega_p + \Omega_\LCm\big(1-\Omega_p^2\big)$, implying
\be\label{Hej-U}
	\Omega_p = \frac{\Omega_\LCp-\pi_0}{\Omega_\LCm} =: \cup_p\;,
\ee
in which the $\Omega_\LCpm$ take constant values after the pulse has turned off, and the given physical solution is that having the correct zero-field limit. (We return to the other solution later.) 
%
%
%
%
{The vacuum persistence amplitude is $\bracket{\mathbb 0}{0;\infty}$, and not $\bracket{0}{0;\infty}$.} Similarly, counting the number of pairs produced after the field has switched off is equivalent to counting free particle excitations in the pure gauge vacuum. The Hamiltonian is easily diagonalised; the normalised operators which create electrons and positrons of spatial momentum~$p$ from $\ket{\mathbb{0}}$ are, respectively, $B^\dagger_{p+a_\infty}$ and $D^\dagger_{p-a_\infty}$, where
\be
	\label{BD-DEF}
	B^\dagger_{p} := \frac{b^\dagger_p - \cup_p d_{-p}}{\sqrt{1+\cup_p^2}}  \;, \qquad D^\dagger_{-p} := \frac{d^\dagger_{-p} + \cup_p b_{p}}{\sqrt{1+\cup_p^2}}  \;.
\ee
The momentum assignments of the modes do not correspond to physical momenta because of the pure gauge terms, but the relation between them is a simple translation.  The number of pairs created from vacuum is equal to the number of created electrons, so the number (density) of pairs of physical momentum $p$ created from a pulse of profile $a(t)$ is given by
\be\label{N-def-1}
	\mathcal{N}(p|a) := V^{-1}\lim_{t\to\infty} \bra{0;t}B^\dagger_{p+a_\infty} B_{p+a_\infty} \ket{0;t} \;. 
\ee
If $E(t)$ has compact support $t < t_f$, we may drop the limit and evaluate (\ref{N-def-1}) at any $t>t_f$.

\section{Adiabatic pair number}\label{sec:adiabatic}
In describing pair production at intermediate times, while the electric field is still turned on, a common choice of basis states is that of adiabatic (or {instantaneous}) eigenstates of the Hamiltonian. The adiabatic vacuum is defined by promoting $\cup_p$ and hence $\ket{\mathbb{0}}$ to time-dependent objects. States are built on the vacuum by the ladder operators~(\ref{BD-DEF}) which similarly become time-dependent. The adiabatic `particle number' $N_p(t)$ is then defined by
$N_p(t) := V^{-1}\bra{0;t} B^\dagger_{p} (t) B_{p}(t)\ket{0;t}$.  By construction we have $\lim_{t\to\infty} N_{p+a_\infty}(t) = \mathcal{N}(p|a)$. Using the explicit form~(\ref{sol1}) the adiabatic number of pairs is
 \be\label{NNN}
	N_p(t) = \frac{|\Omega_p(t) - \cup_p(t)|^2}{(1+|\Omega_p(t)|^2)(1+\cup_p^2(t))} \;.
\ee
For finite time $N_p(t)$ can exhibit `transient' oscillations which are orders of magnitude larger than the final number of pairs; see~\cite{Hebenstreit:2008ae,Kim:2011jw,Blinne:2013via,Zahn:2015awa} {and the figures below.}

$\cup_p(t)$ is the first (lowest order) of an infinite number of approximations to $\Omega_p(t)$ found from performing an adiabatic expansion of the Schr\"odinger equation (\ref{schro}), see~\cite{Dabrowski:2014ica,Dabrowski:2016tsx}. Each order of approximation provides a candidate number of pairs, call it ${}_n N_p(t)$, obtained from (\ref{NNN}) by replacing $\cup_p$ with the $n^\text{th}$ order adiabatic approximation $_n\Omega_p$ of $\Omega_p$; all differ, but yield the same asymptotic $\mathcal{N}$. This and (\ref{NNN}) make explicit an observation in~\cite{Dabrowski:2016tsx}; the number of produced pairs is supported on the difference between the {exact} solution of the Schr\"odinger equation $\Omega_p$ and its adiabatic {approximations} ${}_n\Omega_p$. There is therefore a `pair producing part' of the exact solution to which adiabatic approximations are blind; a trans-series analysis of this result would be interesting to pursue. {Examples of higher-order adiabatic numbers are provided in Appendix~\ref{AppA}, but our focus here is} on finding the meaning of the adiabatic particle number~(\ref{NNN}).

%
%
\subsection{Physical interpretation}
%
Take any given electric field $E(t)$ and imagine instantaneously switching it off at some $t=\tau$. To find the number of created pairs we solve the Schr\"odinger equation for $\Omega_p(t)$ with the continuous background $a_{\tau}(t) := a(t)\theta(\tau-t) + a(\tau)\theta(t-\tau)$.  Note that $a_\tau(\infty) =a_\tau(\tau)$. The solution $\Omega_p(t)$ agrees exactly with that in the background $a(t)$ for $t<\tau$, and crucially is continuous at $t=\tau$. 
For $t>\tau$, $\Omega_p(t)$ obeys the pure gauge version of (\ref{schro}) where $\Omega_\LCpm$ are evaluated at $t=\tau$, together with the boundary condition $\Omega_p(t) = \Omega_p(\tau)$. This part of the solution, at $t>\tau$, is easily found:
\be\label{extra}
	\Omega_p(t) = \frac{\Omega_\LCp-i \pi_0 \tan \big[\pi_0(t-\tau) -i \kappa)\big]}{\Omega_\LCm} \;,
\ee
with $\Omega_\LCpm\equiv \Omega_\LCpm(\tau)$ and $\kappa = \tanh^{-1}[(\Omega_\LCp-\Omega_\LCm \Omega_p(\tau))/\pi_0]$. 
Using (\ref{Hej-U}) and (\ref{extra}) the asymptotic number of pairs is
\be\label{almost}
\begin{split}
	\mathcal{N}(p-a(\tau)|a_\tau)
	&\equiv V^{-1}\bra{0;t} {\hat N}_p(t)\ket{0;t}\Big|_{t\geq \tau} \\
	&= \frac{|\Omega_p(\tau)- \cup_p(\tau)|^2}{(1 + |\Omega_p(\tau)|^2) (1 + \cup^2_p(\tau))} \;,
\end{split}
\ee
constant for $t\geq\tau$ and equal to its value at the switch-off by continuity. The final expression in (\ref{almost}) is nothing but the adiabatic number of pairs (\ref{NNN}) at time $\tau$, calculated in the original field $a(t)$. Hence we have the result
\be\label{Thanos}
	N_p(t) = \mathcal{N}(p-a(t)|a_t) \;.
\ee
{This means that adiabatic number at time $t$ is in fact counting} the number of physical pairs which would be observed, with a shifted momentum, if the field were snapped off at time~$t$. The remainder of the paper is devoted to analysis and discussion of this result.

\begin{figure}[t!]
\includegraphics[width=0.9\columnwidth]{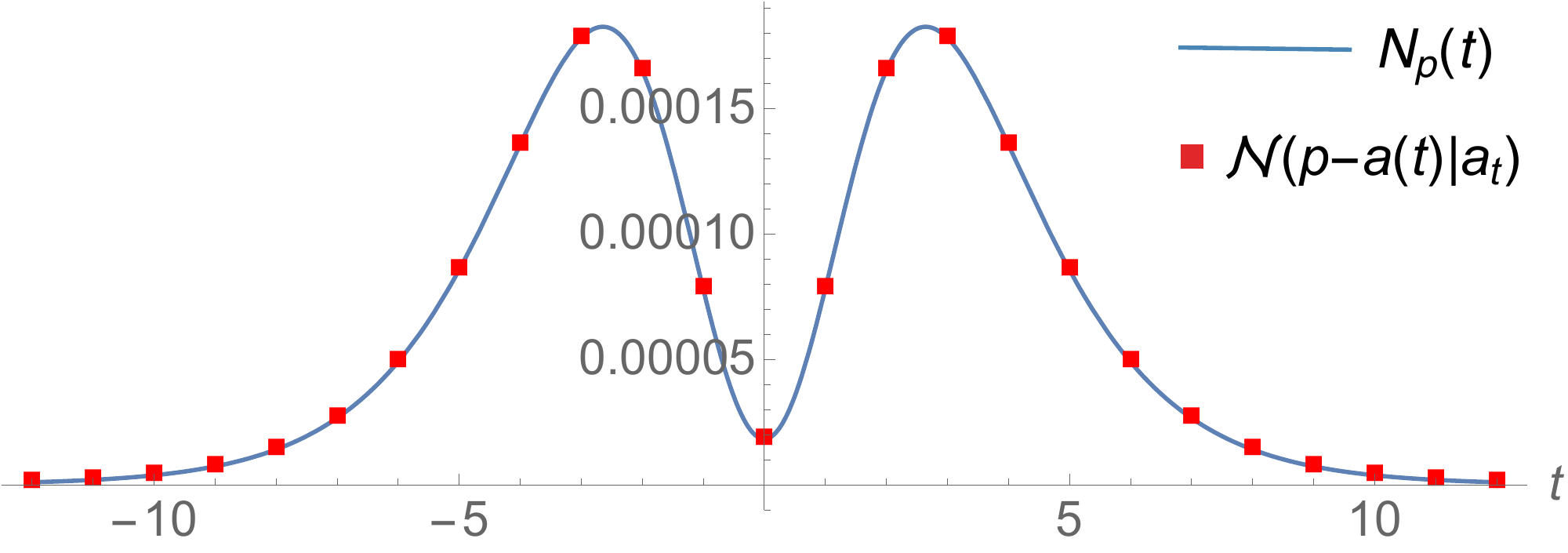}
\includegraphics[width=0.9\columnwidth]{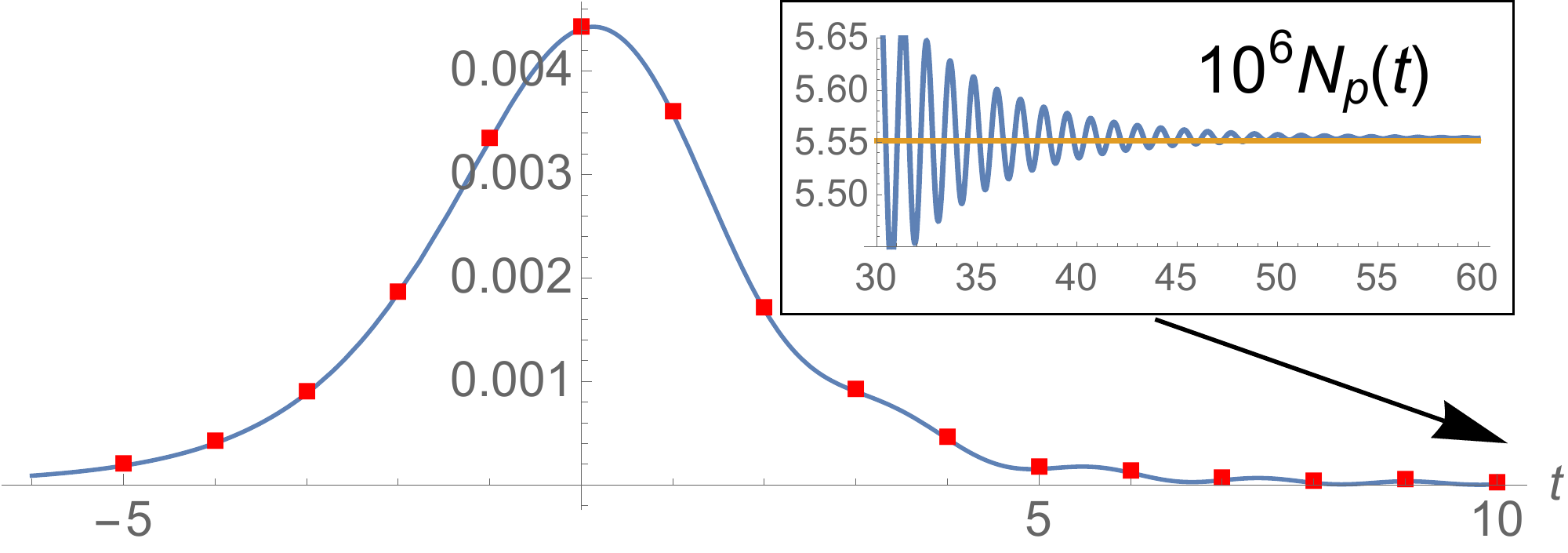}
\caption{\label{FIG:RESULTAT} {\emph{Upper panel:} Comparison of adiabatic particle number $N_p(t)$ and asymptotic number $\mathcal{N}(p-a(t)|a_t)$ for $a(t)=1/3\, \text{sech }(t/3)$ and $p=0$.  $N_{p=0}(t)$ is calculated analytically from $\Omega_{p=0}$ in the text, while $\mathcal{N}(p-a(t)|a_t)$ is calculated numerically; the results are identical, verifying~(\ref{Thanos}). (All in units where $m=1$.) {\emph{Lower panel}: the same comparison for a Sauter pulse $a(t) = eE_0/\omega (1+\text{tanh}\,\omega t)$, with $E_0=1/4$, $\omega=1/10$ and $p=5/2$. $N_p(t)$ contains a great deal of structure and the peak at $t\simeq0$ is three orders of magnitude larger than the final number of pairs at $t\to\infty$.}}}
\end{figure}

%
The result (\ref{Thanos}) may be verified by calculating $\mathcal{N}(p-a(t)| a_t)$ in the field $a_t$ and comparing against $N_p(t)$ in the field~$a$. {This is shown in Fig.~\ref{FIG:RESULTAT} for a Sauter pulse and for
the field} $a(t) := (1/ \lambda) \text{sech}\, t/\lambda$ and $p=0$,
for which the exact solution to the Schr\"odinger equation is $\Omega_{p=0}(t) = (2\lambda m\cosh t/\lambda + i \sinh t/\lambda)^{-1}$. This is an example of a `solitonic' pair of field and momentum for which $N_0(t)\not=0$ but $\mathcal{N}(0|a)= 0$ and there is no pair production asymptotically~\cite{Kim:2011jw,Kim:2011sf,Kim:2012azq,Huet:2014mta,Cai:2019vow}. However, if we turn the field off at any time $t$, the solitonic property is lost and there are pairs.

Other methods could have been used to derive the results above, e.g.~Bogoliubov transforms~\cite{Kim:2008yt,Kim:2009pg} or kinetic equations. {We emphasise, though, that the same choice of basis and ambiguities arise in kinetic approaches, as demonstrated by in~\cite{Kim:2011jw,Huet:2014mta}}. Without a good understanding of the physics implied by the basis choice, one can obtain unphysical results -- such a problem was encountered {in~\cite{Kim:2011jw}, and we solve this using our first-principles approach in Appendix~\ref{AppC}.

\subsection{Smooth switch-off and pulse shaping}
%
{What we have established so far is that the adiabatic basis is not simply a mathematical artefact, but has an interpretation which is, essentially, physically sensible. {The steep field gradient of a sudden turn-off naturally adds higher frequency modes to the field, which assist the production of pairs, but it is obviously not experimentally realisable; we therefore turn now more toward phenomenology, and ask the question:} to what extent do our results hold in the more physical case that the electric field turns off rapidly and smoothly, rather than instantaneously? To address this, define}  $a_{\tau,\Delta}(t)$ by replacing the step functions in $a_\tau(t)$ with smooth functions, such that the larger $\Delta$ is, the more rapidly the field switches off. Fig.~\ref{FIG:MJUKT} gives two examples of smoothing function applied to the Sauter pulse, and of $N_p(t)$. If $\Delta$ is large then the numbers of pairs is, as shown, not greatly affected by using a smooth switch-off, and $\mathcal{N}(p-a_t(t) |a_{t,\Delta}) \simeq \mathcal{N}(p-a_t(t)|a_{t})$.  As $\Delta$ becomes smaller, the asymptotic number of pairs $\mathcal{N}(p-a_t(t) |a_{t,\Delta})$ may be larger or smaller than the adiabatic number, depending on the smoothing used but, importantly, it continues to track the adiabatic number, independent of the choice of smoothing function. This includes both the large peaks of $N_p(t)$ and its finer details, see Fig.~\ref{FIG:MJUKT}. 

\begin{figure}[t!]
\includegraphics[width=0.85\columnwidth]{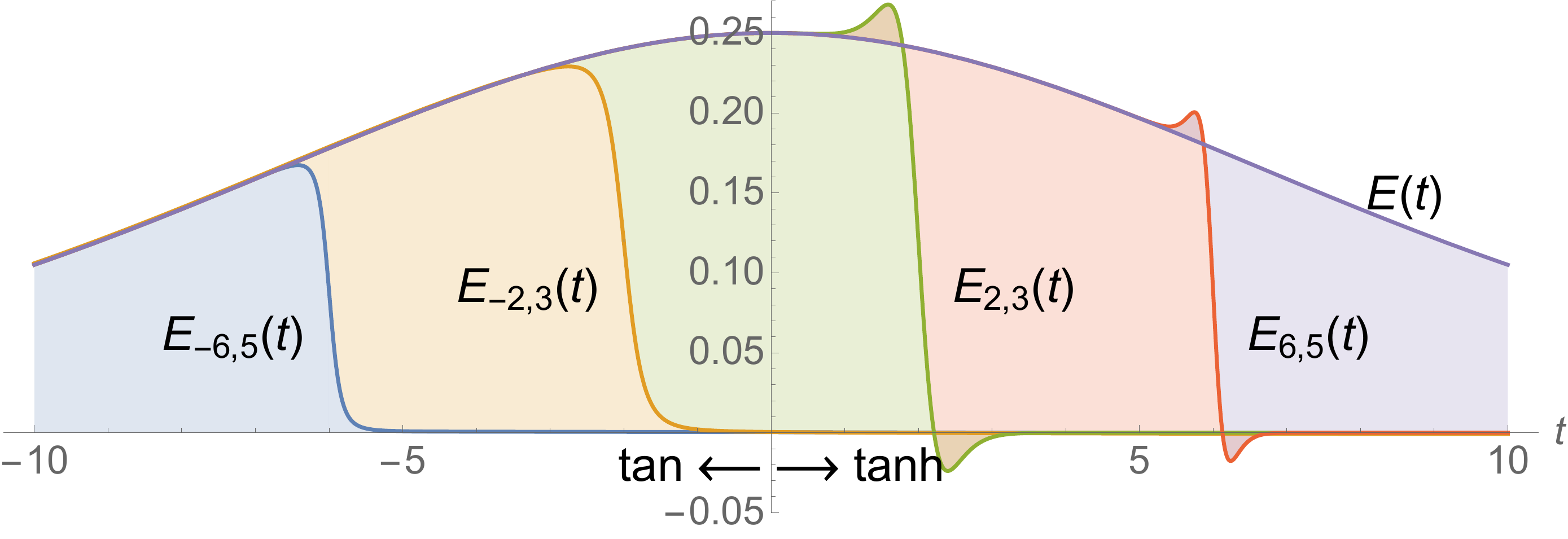}
\includegraphics[width=0.85\columnwidth]{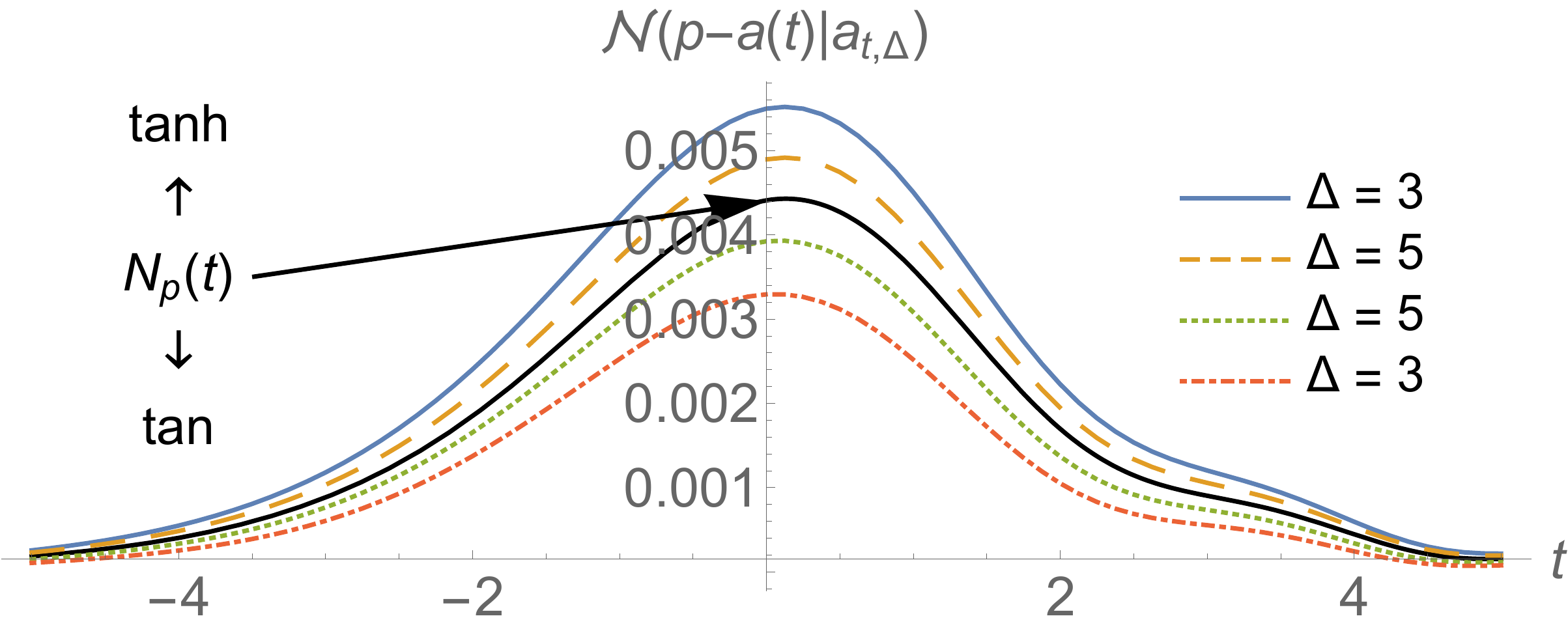}
\includegraphics[width=0.85\columnwidth]{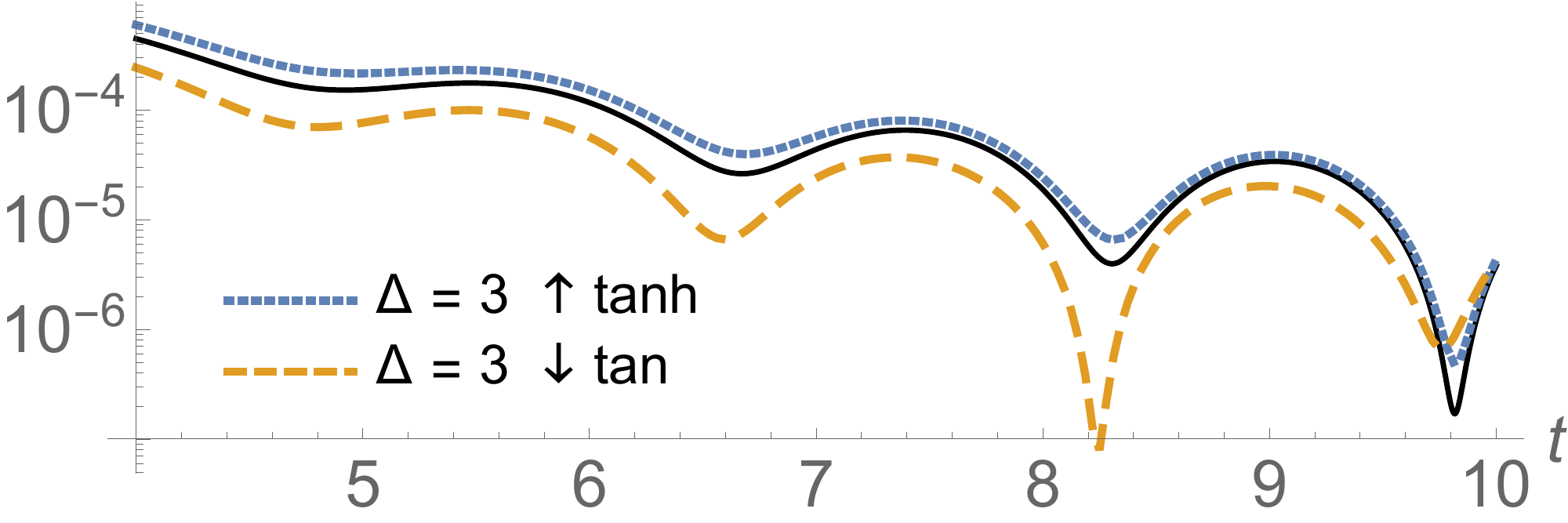}
\caption{\label{FIG:MJUKT} \textit{Top}: Electric fields $E_{T,\Delta}(t)$ with smooth turn off defined by two step regularisations; $\theta (t) \to \tfrac12(1 + \tanh \Delta t )$ and $\theta(t)\to \tfrac12 + \tfrac1\pi \tan^{-1}\Delta t$. 
\textit{Middle}: Asymptotic particle number $\mathcal{N}(p-a(t)|a_{t,\Delta})$ for different smoothing functions and parameters (colour), compared to adiabatic number $N_p(t)$(black). {Electric field and momentum parameters are as for the Sauter pulse in Fig.~\ref{FIG:RESULTAT}.} The asymptotic number tracks the adiabatic, including its large peaks and (\textit{bottom}) finer structure.}
\end{figure}

If the turnoff is slow (small $\Delta$), such that the shape of the electric field is changed significantly to the past of the switch-off time, then the asymptotic and instantaneous numbers obviously differ -- this is also consistent with the non-Markovian nature of pair production~\cite{Rau:1994ee,Kluger:1998bm}, in that the process depends on its past history. 

As the final number of pairs tracks the adiabatic number even with a smooth (but rapid) turn off, it {confirms} that one can in principle, meaning \textit{with sufficiently good pulse shaping}, produce numbers of pairs similar to that predicted by the adiabatic $N_p(t)$. Crucially, this means that large pair numbers seen at intermediate times, such as in Fig.~\ref{FIG:RESULTAT}, are not something to be removed or avoided, but can in principle be pursued. The first step is to identify where the large peaks in $N_p(t)$ lie. Using the Schr\"odinger equation to simplify $\partial_t N_p(t)$, one finds that there are always extrema given by
%
%
%
$\partial_t\cup_p(t)=0$, i.e.~$E(t)=0$, which are not of interest. Writing $\Omega_p(t) = x(t) +i y(t)$, one finds that non-trivial extrema lie on a circle in the complex plane,
\be\label{villk2}
	(x- \Omega_\LCp/\Omega_\LCm)^2 + y^2 = \pi_0^2 /\Omega_\LCm^2 \;,
\ee
which, once  $\Omega_p(t)$ is known, is to be solved for $t$ to find the local maxima of $N_p(t)$. Note that if the imaginary part is zero then (\ref{villk2}) reduces to $\Omega_p=\cup_p$, implying that $\ket{0;t}$ has collapsed back to the vacuum. Hence we are only interested in solutions of (\ref{villk2}) with, somewhat naturally, an imaginary part.  To illustrate, we can solve (\ref{villk2}) in the solitonic case, using the exact form of $\Omega_{p=0}(t)$; the maxima occur at $t = \lambda \log(\sqrt{2}\pm 1)$. (That these are the maxima of $E(t)$ seems to be a coincidence of this case.)

\begin{figure}[t!]
\includegraphics[width=0.95\columnwidth]{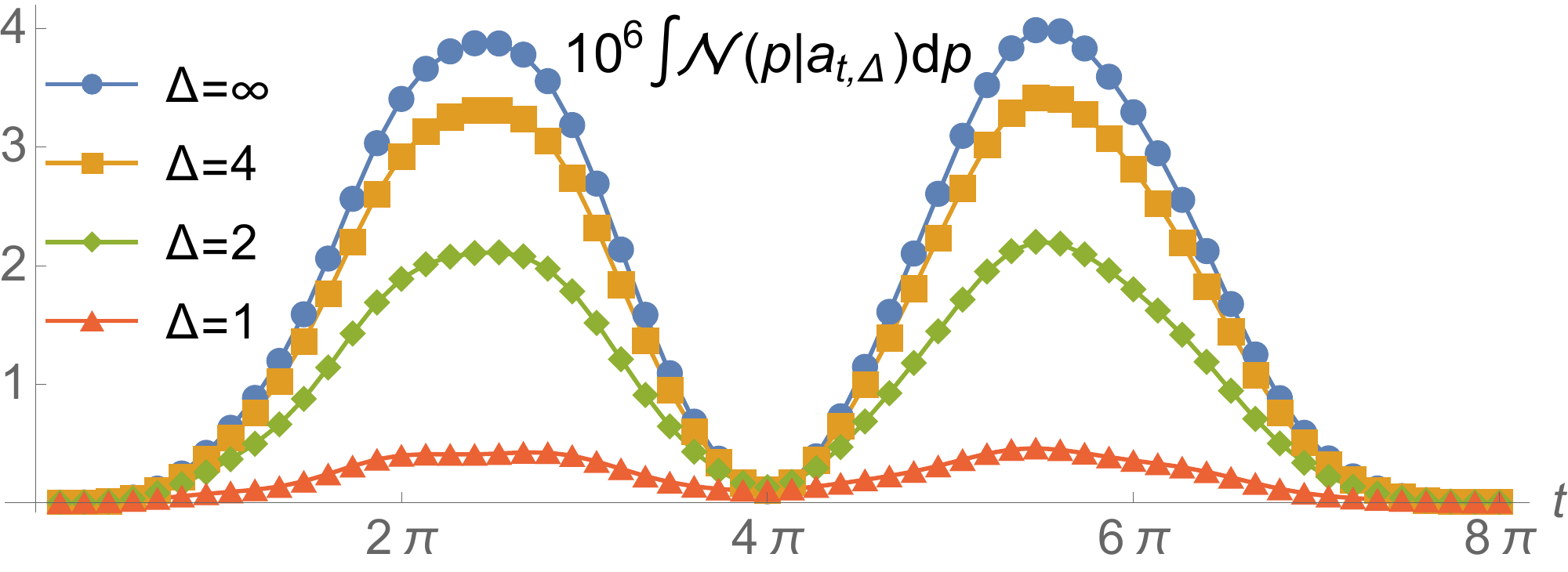}
\caption{\label{FIG:NYTT} {Total pair yield in the electric field
$a'(t) = E_0\partial_t \sin^3(\omega t)$ for {a weak field $E_0=1/20$} and $\omega=1/8$, switched off using the regulated step function $1/2\text{erfc}(\Delta(t-t_f))$. The adiabatic number/hard cutoff (\ref{nyekvation}) is obtained for $\Delta\to\infty$ (top curve). Compared to this, a slow turn off ($\Delta=1$, lowest curve) reduces the total yield, but even this yield remains orders of magnitude higher than the total yield in the original pulse, which in this example is $\sim 10^{-6}$.}}
\end{figure}

\subsection{{UV behaviour and total yield}}

We remark that, {in expanding universes, the total adiabatic particle yield is UV divergent, and requires higher-order adiabatic terms for its regularisation~\cite{Parker:1969au,Fulling:1979ac,Parker:2012at}.  If the same were true in QED, it could cast doubt on the interpretation of (\ref{Thanos}). We show in Appendix~\ref{AppB}, though, by repeating our calculations in 1+3 dimensions to capture the correct UV behaviour, that the total yield in QED is indeed finite.}

With this, {a natural question is whether the preceding results for larger pair yields at fixed momentum translate into larger \emph{total} pair yields. To answer this in the most direct manner, we make a simplification by calculating $N_p(t)$ perturbatively, to lowest order in powers of the field. Using (\ref{NNN}) and (\ref{schro}) we find} 
\be\label{nyekvation}
	{\int\!\ud p\, N_p(t) \simeq \int\!\ud p\, \frac{m^2}{4p_0^4} \bigg|\int\limits_{-\infty}^t\!\ud s\, a'(s)e^{2ip_0s}\bigg|^2} \;.
\ee
We calculate this total adiabatic yield in Fig.~\ref{FIG:NYTT} {for a weak electric field with a $\sin^3$ profile}, and again compare with (asymptotic) yields for which we include a smooth switch-off. We note two results. First, the total yield, like the differential, can be higher than in the original field. Second, while a slow {and therefore more physical} turn off gives a smaller yield than the adiabatic result, it is still significantly higher than the original yield. {The parameters chosen in Fig.~\ref{FIG:NYTT} are not special; the field strength, for example, is clearly an overall scaling in this perturbative approximation. We only want to demonstrate the principle that the total yield can be enlarged. The extent to which such increases can be achieved in realistic experimental scenarios is beyond the scope of this paper.}

%

%

{We comment briefly on why the number of pairs is not monotonic increasing in time. Pairs created at different times with different momenta can be accelerated by the field to the same asymptotic momentum -- this leads to quantum (path) interference which can enhance or deplete the pair number, see~\cite{King:2010nka,Hebenstreit:2009km,Akkermans:2011yn,Ilderton:2019ceq} for examples. Created pairs can also annihilate into the field via the `inverse' Schwinger process; this is not often discussed in the literature, but see~\cite{PhysRevA.74.044103,Nishida:2021qta}, so we provide example calculations in Appendix~\ref{AppD} which show that the annihilation probability can be non-zero. 
}

\section{Discussion}\label{sec:outro}
We have shown that adiabatic particle number in the Schwinger effect has a definite physical interpretation. The adiabatic number of pairs at any given time is the number of physical pairs which would be observed, with a properly identified momentum, if the field were (very) rapidly switched off at that time. As a lesson in pulse-shaping, our results say that steep field gradients (implying high frequency components) can be beneficial for creating the very large number of pairs predicted by adiabatic particle number.

Our results suggest a change in perspective: rather than trying to identify a `correct' pair number, one can instead try to identify what a given number operator really measures,  as different bases describe different observables. The challenge is to understand what these are. As a simple first example in this  line of investigation, we recall from the quadratic above (\ref{Hej-U}) that there is a second solution of the Schr\"odinger equation describing the (pure gauge and) adiabatic vacuum. This solution, $\cap_p :=  -1/\cup_p$, diverges in the free-field limit and so will not count the asymptotic number of pairs correctly, but when the field is on we can still define a vacuum and basis of states from it, and an instantaneous number of excitations (of something) ${\tilde N}_p(t)$ as in (\ref{NNN}) but with $\cup_p$ replaced by $\cap_p$.  The physical content of this basis is easily found: a direct calculation shows that ${\tilde N}_p(t)$ is, due to Pauli blocking,  the `unoccupied' number density ${\tilde N}_p(t) = 1- N_p(t)$, i.e.~${\tilde N}_p$ is trivially counting one-minus the adiabatic number of pairs, for which we have already established the physical meaning.


There are several topics for future research. {One is to identify the physical meaning of higher-order and super-adiabatic bases, and in this context explore the properties of operators such as the current~\cite{Zahn:2015awa,Aleksandrov:2020mez,Aleksandrov:2021ylw}.  Another is to pursue the impact of pulse shaping {in more realistic settings which include e.g.~spatial inhomogeneities, and to} identify optimal field configurations for Schwinger pair production~\cite{Blaschke:2012vf,Kohlfurst:2012rb,Linder:2015vta,Dong:2017vse,Fillion-Gourdeau:2017uss,Aleksandrov:2017owa,Lv:2018sqm,Unger:2019zjb,Unger:2019lhp}.} The results here can also be extended to related, and analogue, Schwinger effects in other areas of physics, such as monopole production~\cite{Gould:2019myj}.

\begin{acknowledgments}
\textit{A.I.~thanks B.~King for comments on the manuscript, and G.~Torgrimsson for many useful, fruitful discussions.}
\end{acknowledgments}

\appendix
\onecolumngrid

%
\section{(Super) adiabatic bases in the Schr\"odinger picture}\label{AppA}
%
Though normally considered via solutions of the Klein-Gordon and Dirac equations~\cite{Dabrowski:2014ica,Dabrowski:2016tsx}, the adiabatic expansion of the Schr\"odinger equation is easily found; we rearrange (2) in the text to write the `solution' as
\be\label{omordnat}
	\Omega_p(t) = \frac{\Omega_\LCp(t) - \sqrt{\pi_0^2(t) - i \hbar\Omega_\LCm(t) \partial_t\Omega_p(t)}}{\Omega_\LCm(t)} \;,
\ee
in which we have re-instated $\hbar$ and fixed the sign of the square root using the zero-field limit. The adiabatic expansion is an expansion in powers of time derivatives, which are clearly counted here simply by powers of $\hbar$. We denote the $n^\text{th}$ order adiabatic expansion of $\Omega_p$ by $_n\Omega_p$; low orders are easily constructed from (\ref{omordnat}),
 \be
	_0\Omega_p(t) =  \frac{\Omega_\LCp(t) - \pi_0(t)}{\Omega_\LCm(t)} \equiv \cup_p(t) \;, \qquad _1\Omega_p(t) = \cup_p(t)  +i \hbar \frac{\partial_t\cup_p(t)}{2 \pi_0(t)} \;,
 \ee
although general expressions for higher $n$ quickly become unwieldy. (We note in passing that one could perform a partial resummation of the series by retaining the square root in (\ref{omordnat}) and expanding only under it; for the examples considered here the effect seems to be roughly equivalent to advancing the entirely perturbative approach by one order.) We can then define an $n^\text{th}$ order adiabatic number of pairs ${}_n N_p$ from the `vacuum' ${}_n\Omega_p$ and excitations built on it using operators like those in (4) in the text.  Provided we deal with electric fields which vanish smoothly as $t\to\infty$, each of the $_n\Omega_p$ go over to $_0\Omega_p = \cup_p$ asymptotically, and so each give the same value for the asymptotic number of pairs. We have:   
\be\begin{split}\label{O-minus-U-allm}
{_n} N_p(t) := \frac{|\Omega_{p}- {_n}\Omega_{p}|^2}{(1+|\Omega_p|^2)(1+|{_n}\Omega_{p}|^2)} \;, \qquad\quad  \mathcal{N}(p|a)  =  \lim_{t\to\infty}  {_n} N_{p+a_\infty}
	 \quad \text{for all  }   n \;.   
\end{split}\ee
We illustrate this in Fig.~\ref{FIG:SUPER} using the Sauter pulse 
\be\label{Exempel-till-fig-super}
	E(t) = E_0 \text{sech}^2(\omega t)  \;, \qquad a(t) = \frac{eE_0}{\omega} \big(1+\text{tanh}(\omega t)\big) \;,
\ee
and see immediately the appearance of the super-adiabatic behaviour presented in~\cite{Dabrowski:2014ica,Dabrowski:2016tsx}; as the order $n$ of the adiabatic expansion increases, the transient oscillations in the $_nN_p(t)$ decrease in amplitude to an almost monotonically increasing function interpolating between $0$ and the asymptotic value $\mathcal{N}$. This continues until an `optimal' order, here $n=6$, beyond which (we have confirmed) the oscillations re-emerge. From (\ref{O-minus-U-allm}) we see that pair production is supported entirely on the difference between $\Omega_p(t)$ and its adiabatic expansions.  Similarly to~\cite{Dabrowski:2016tsx}, we find that there is a phase difference between the real and imaginary parts of $(\Omega_p - {}_n\Omega_p)^2$  in the numerator of ${}_n N_p(t)$ which is responsible for the oscillations in the pair number, and that this sums to give an almost monotonic, smooth function at the optimal truncation order. The physical interpretation of these higher-order adiabatic particle numbers (also in cosmology~\cite{Yamada:2021kqw}) is an intriguing topic for future work.

\begin{figure}[t!]
\includegraphics[width=0.32\textwidth]{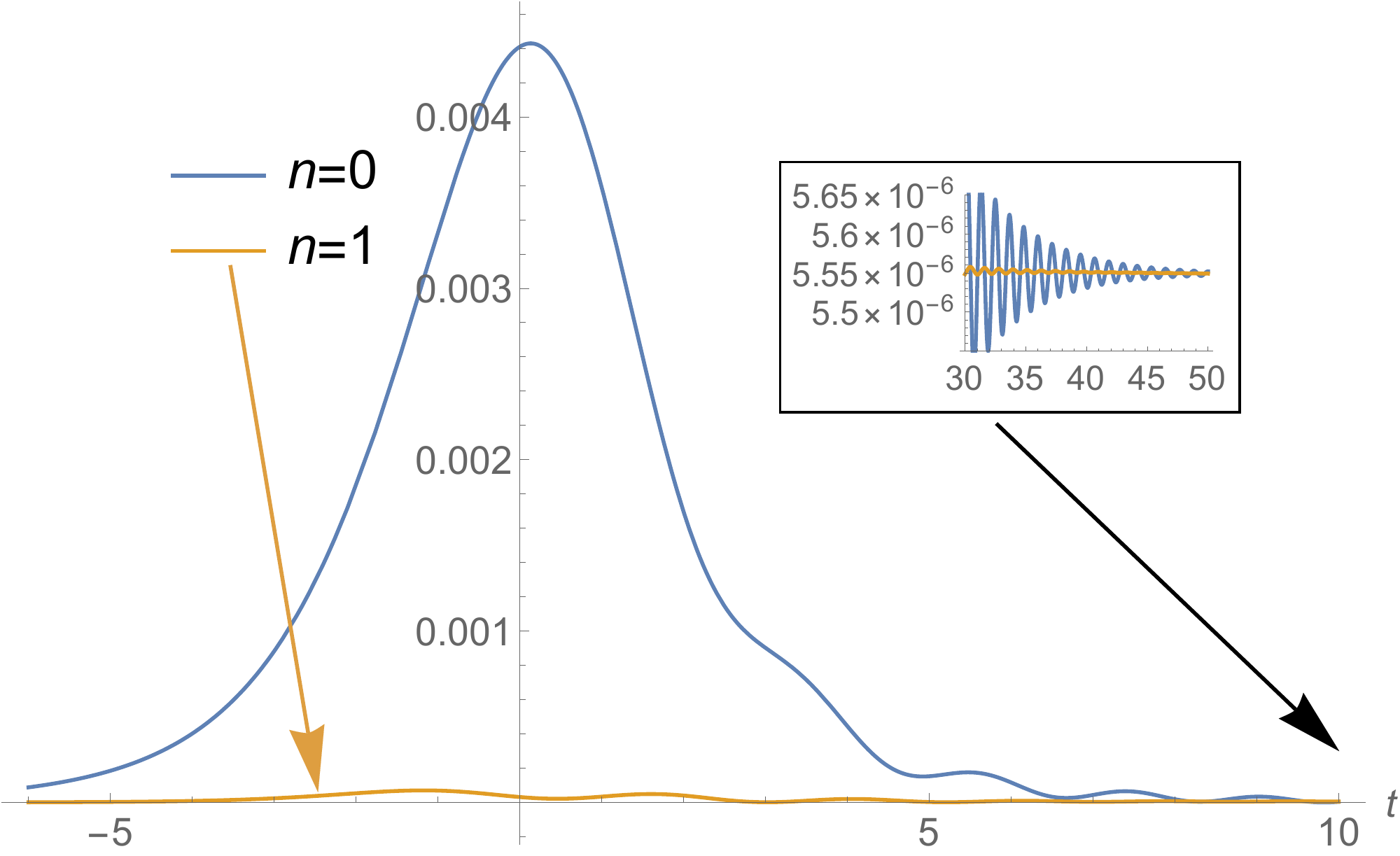}
\includegraphics[width=0.32\textwidth]{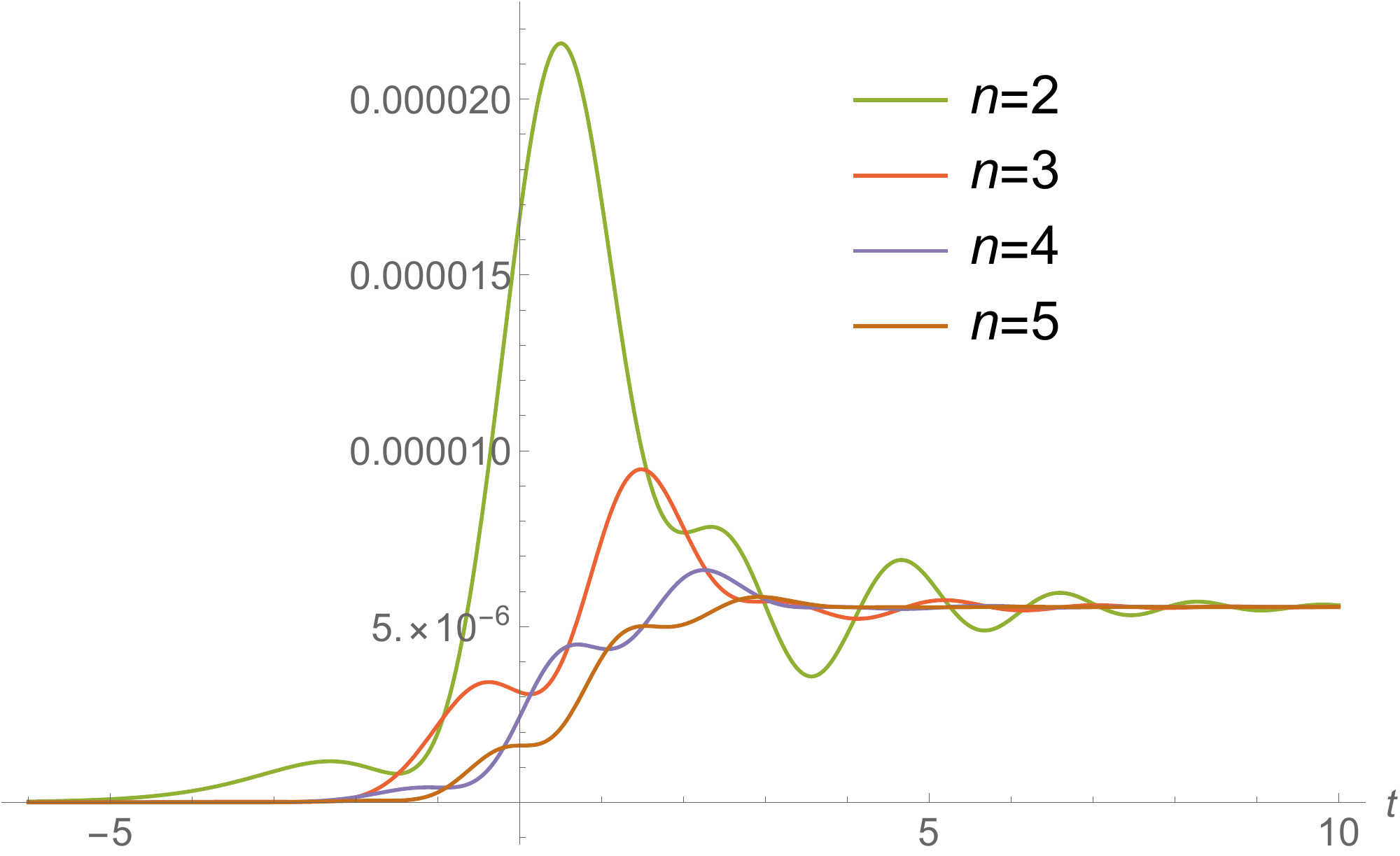}
\includegraphics[width=0.32\textwidth]{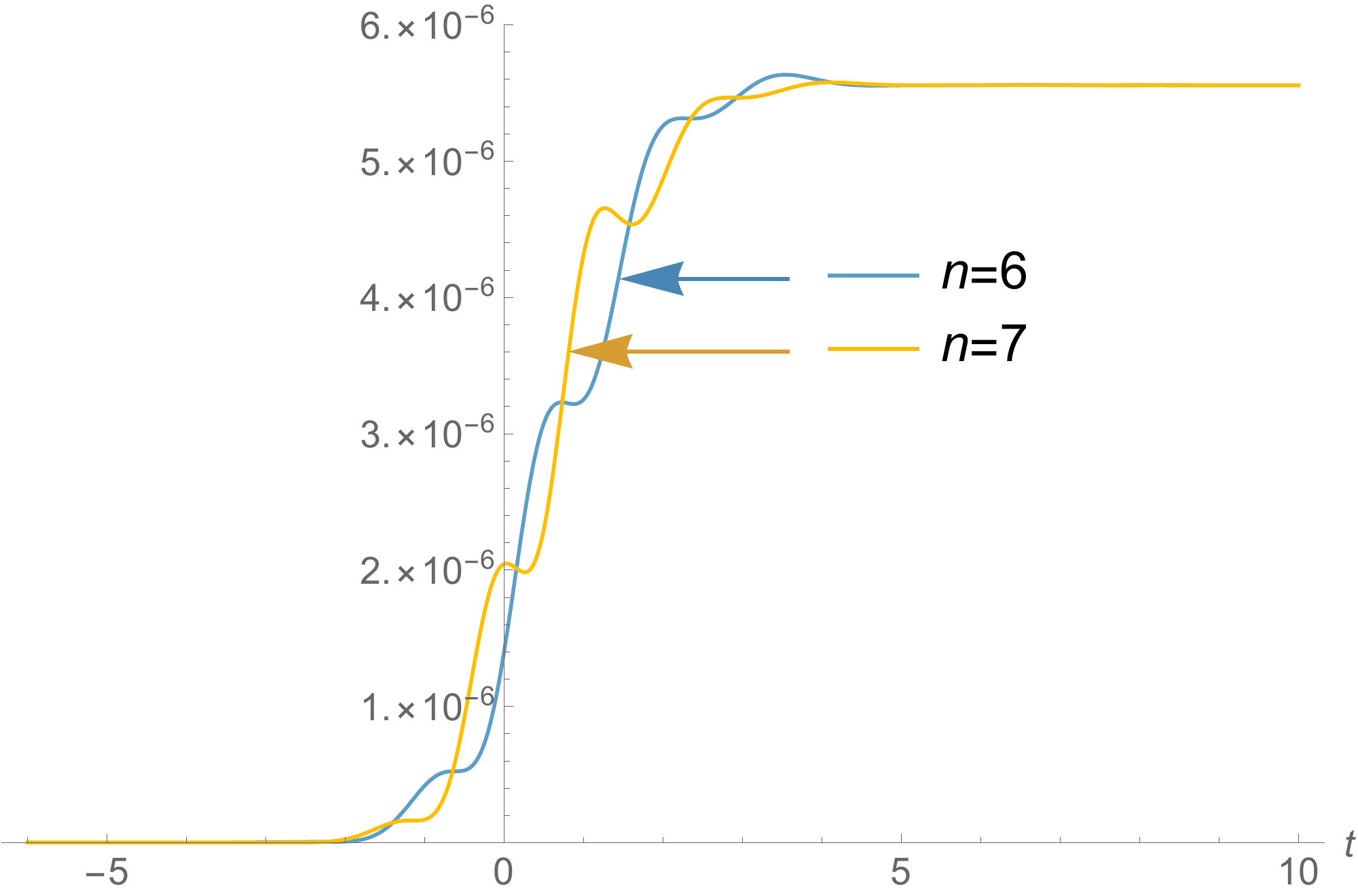}
\caption{\label{FIG:SUPER} (Super)-adiabatic particle number in the Sauter pulse (\ref{Exempel-till-fig-super}) with parameters $E_0=1/4$, $\omega=1/10$ and $p=5/2$. The fluctuations in particle number decrease in size as the order $n$ of the adiabatic expansion increases, up to $n=6$, then begin to re-emerge.}
\end{figure}

\section{Scalar pair production in 3+1, UV behaviour \& benchmarking}\label{AppB}
An objection to the interpretation of our results could be raised by observing that, in the analogous process of particle creation from an expanding universe, the lowest order adiabatic particle number is, when summed over all momenta, UV divergent~\cite{Parker:1969au}. This is unphysical and, moreover, in order to define a finite number of created pairs one must include higher order terms in the adiabatic expansion~\cite{Parker:2012at}.

As UV behaviour is sensitive to the number of dimensions, we give here the extension of our methods and results to 1+3 dimensions. We will show that the total number of pairs created according to our results is UV finite. For this investigation we go to scalar rather than spinor QED: this is partly for simplicity, and partly as it will afford us an opportunity to benchmark against the literature. There are many similarities to QED 1+1, so we can be brief.

The gauge field is now $eA_\mu= (0,0,0,a(t))$ and the Hamiltonian is, for $\phi$ the charged scalar and $\Pi$ its canonical momentum,
\be
	H(t) = \int\!\ud^3\mathbf{x}\,\, \Pi^\dagger \Pi 
	+ |\partial_\LCperp \phi|^2
	+ |i \partial_z \phi - a(t)\phi|^2 
	+m^2|\phi|^2\;.
\ee
We use the usual free-field mode expansions,
\be
	\phi = \int\!\frac{\ud^3{\bf p}}{\sqrt{(2\pi)^22p_0}}\,\, e^{-i{\bf p}.{\bf x}}\big( a_{\bf p}   + b^\dagger_{-\bf p}\big) \;, 
	\qquad
	\Pi = \int\!\frac{\ud^3{\bf p}}{\sqrt{(2\pi)^22p_0}}\,\, i p_0 e^{i{\bf p}.{\bf x}}\big( a_{\bf p}   - b^\dagger_{-\bf p}\big) \;, 
\ee
with $[a_{\bf p},a_{\bf q}^\dagger] = [b_{\bf p},b_{\bf q}^\dagger] = \delta^3({\bf p}-{\bf q})$. The Hamiltonian takes a similar form to that in QED 1+1,
\be
	H(t) = \int\!\ud^3\mathbf{p} \, \Omega_\LCp(t) (a^\dagger_{\bf p} a_{\bf p}
 + b^\dagger_{-\bf p} b_{-\bf p}) + \Omega_\LCm(t) (a^\dagger_{\bf p} b^\dagger_{-\bf p}+a_{\bf p} b_{-\bf p}) \;,
 \ee
but note that the definitions of $\Omega_\LCpm$ differ:
\be
\qquad \Omega_\LCpm(t) = \frac{\pi_0^2(t) \pm p_0^2}{2p_0} \;,
\ee
in which the classical energy is now $\pi_0:=\sqrt{p_\LCperp^2 + (p_z-a(t))^2+m^2}$, with $p_\LCperp =(p_x,p_y)$. The time-evolved vacuum state is
\be\label{scalar-schro}
	\ket{0;t} = \exp\bigg[-i V \vartheta(t)  - \int\!\ud^3{\bf p}\, \Omega_{\bf p}(t) a^\dagger_{\bf p}b^\dagger_{-\bf p}\bigg] \ket{0;t} \;,
	\qquad
	 i \partial_t {\Omega}_{\bf p} = 2 \Omega_\LCp \Omega_{\bf p} - \Omega_\LCm (1 + \Omega_{\bf p}^2) \;,
\ee
in which the Schr\"odinger equation for the covariance $\Omega_{\bf p}(t)$, shown on the right, differs in signs compared to that for spinors. From these expressions we read off the adiabatic vacuum and associated creation/annihilation operators:
\be\label{skalar-vak}
	\cup_{\bf p}(t) = \frac{\Omega_\LCp(t) - \pi_0(t)}{\Omega_\LCm(t)} \;, \qquad A^\dagger_{\bf p} = \frac{a^\dagger_{\bf p} + \cup_{\bf p} b_{-\bf p}}{\sqrt{1-\cup_{\bf p}^2}} \;, \qquad
	B^\dagger_{-\bf p} = \frac{b^\dagger_{-\bf p} + \cup_{\bf p} a_{\bf p}}{\sqrt{1-\cup_{\bf p}^2}} \;.
\ee
The adiabatic number of pairs $N_{\bf p}(t)$ is easily found. Solving the Schr\"odinger equation for a field which is snapped off at some time $t$, one finds the same direct interpretation of adiabatic pair number as for 1+1 QED, (9) in the text:
\be\label{SKALAR-N}
	N_{\bf p}(t) = \frac{|\Omega_{\bf p}(t) - \cup_{\bf p}(t)|^2}{(1-\cup_{\bf p}^2(t))(1-|\Omega_{\bf p}(t)|^2)} = \mathcal{N}(p_\LCperp,p_z-a(t)|a_t) \;.
\ee
We pause to benchmark our expression for $N_{\bf p}(t)$ against~\cite{Dabrowski:2014ica}, in which more common methods based on `second order' approaches were used. The reason for doing so is that, if the adiabatic particle number really is a good observable as we claim, then we should be able to calculate using any method we choose. Fig.~\ref{FIG:SCALAR-GD} shows exact agreement with~\cite{Dabrowski:2014ica}.


%
The total number of pairs created in the adiabatic approximation is given by integrating $N_{\bf p}(t)$ over all momenta ${\bf p}$. We will show that this integral is UV finite. To do so we need to know how $N_{\bf p}(t)$ behaves in the UV, i.e.~for large $|{\bf p}|$. Going to polar coordinates, let $p_z = |{\bf p}|\cos(\theta)$ and $|p_\LCperp| = |{\bf p}|\sin(\theta)$. We now solve the Schr\"odinger equation (\ref{scalar-schro}) for an arbitrary background $a(t)$ in a large $|{\bf p}|$ expansion. We find
\be\label{Stort-P-1}
	\Omega_{\bf p}(t) = -\frac{a(t) \cos(\theta)}{2 |{\bf p}|}-\frac{a(t)^2 \cos (2\theta)+i a'(t)\cos (\theta)}{4 |{\bf p}|^2} + \mathcal{O}(|{\bf p}|^{-3}) \;.
\ee
The corresponding expansion of the adiabatic vacuum (\ref{skalar-vak}) is
\be\label{Stort-P-2}
	\cup_{\bf p}(t) = -\frac{a(t) \cos (\theta)}{2 |{\bf p}|} - \frac{a(t)^2\cos(2\theta)}{4 |{\bf p}|^2} + \mathcal{O}(|{\bf p}|^{-3}) \;.
\ee
The leading order terms, in powers of $|{\bf p}|^{-1}$, in (\ref{Stort-P-1}) and (\ref{Stort-P-2}) are the same, hence it is necessary to go to next to leading order to identify the first nonzero contribution to the number of pairs. Interestingly, the real part of the next to leading order term in (\ref{Stort-P-1}) is also exactly equal to that in the adiabatic vacuum. There is, though, an imaginary part on which the number of pairs has its support, and we conclude from (\ref{SKALAR-N})--(\ref{Stort-P-2}) that
\be
	N_{\bf p}(t) \sim \frac{{a'(t)}^2}{16 |{\bf p}|^4}\cos^2\theta  + \mathcal{O}(|{\bf p}|^{-5})\;.
\ee
Dropping irrelevant constants, the UV contribution to the total number of pairs thus behaves as
\be
	\int\!\ud^3{\bf p}\; N_{\bf p}(t) \sim a'(t)^2\int\limits^\infty\! \frac{\ud |{\bf p}|}{|{\bf p}|^2}  \;,
\ee
and is finite. This supports our interpretation of the adiabatic particle number as an observable.

It would, in this light, be very interesting to reconsider the properties of operators other than the number of pairs, such as the current, its adiabatic approximations, and their renormalisation~\cite{Kluger:1992gb,Zahn:2015awa,Beltran-Palau:2020hdr}.

\begin{figure}[t!]
	\includegraphics[width=0.4\textwidth]{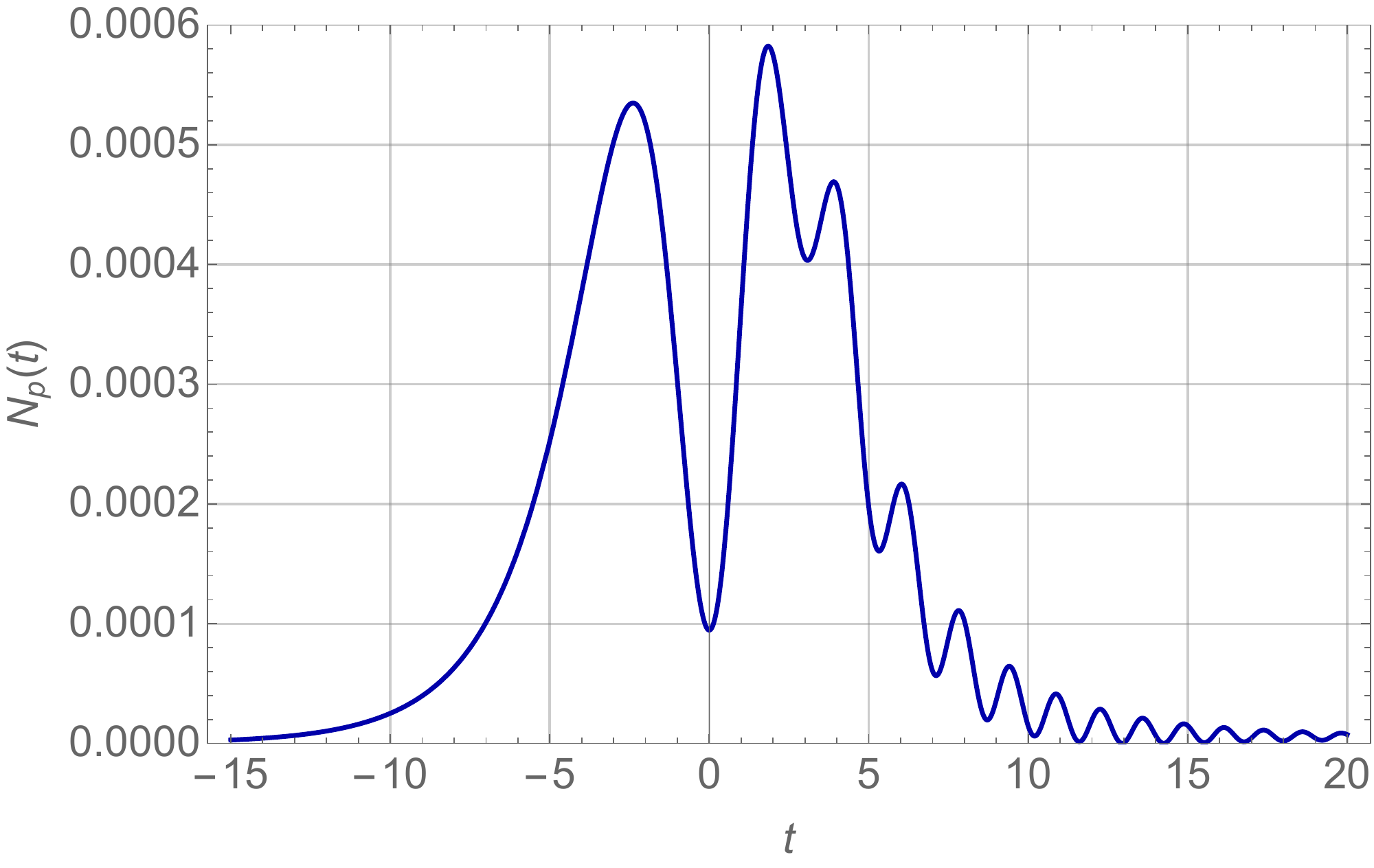}
	\qquad
	\includegraphics[width=0.4\textwidth]{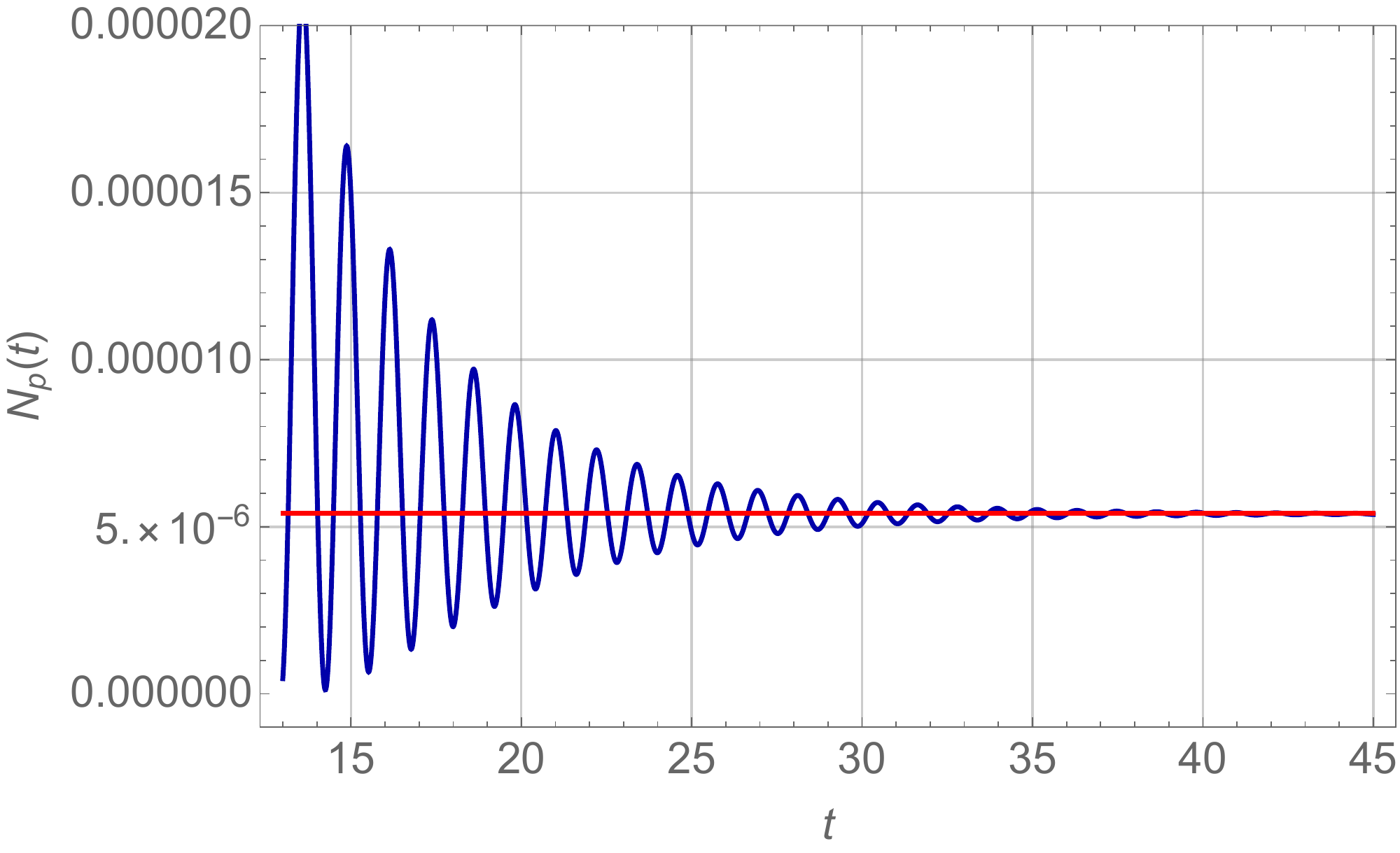}
	\caption{
	\label{FIG:SCALAR-GD}
	Adiabatic pair number (\ref{SKALAR-N}) in the Sauter pulse $E_z(t) = E_0 \text{sech}^2\omega t$, $a(t) = eE_0/\omega (1+\text{tanh}\omega t)$, with $E_0=1/4$, $\omega=1/10$, $p_\LCperp=0$ and $p_z=5/2$, corresponding to final physical $z$-momentum $= -5/2$, in units where $m=1$. Parameters are chosen to benchmark against Fig.~1 in~\cite{Dabrowski:2014ica}; our physical momentum assignments correspond to $k_\LCperp = 0$ and $k_\parallel=0$ in that paper (where the gauge potential is not zero in the asymptotic past): the curves are identical to those in \cite{Dabrowski:2014ica} and we find the same asymptotic number of pairs, as highlighted by the red line in the second panel.
}
\end{figure}

\section{Pair production in a delta-function pulse}\label{AppC}
We return to 1+1 dimensions, considering here pair production in a delta-function electric field, $E(t)\sim \delta(t)$ meaning a potential step $a(t) := a_\infty \theta(t)$ with $a_\infty$ a constant. The potential is everywhere pure gauge, except at $t=0$. This setup was considered in~\cite{Kim:2011jw} using kinetic equations, but observables such as the number of produced pairs were found to carry an unphysical time dependence even for $t\gg 0$, where the field is wholly absent.  We stress that the resolution of this issue can be deduced within kinetic theory from~\cite{Huet:2014mta}, but it does not seem to have appeared explicitly in the literature. We present it here. This section helps to emphasise that understanding the choice of basis clarifies the physics of pair creation.

As in the text, we need to solve the Schr\"odinger equation to identify how the vacuum evolves through the delta-function pulse; the equation for the vacuum state covariance $\Omega_p(t)$ is
\be\label{delta-schro}
	i \partial_t \Omega_p(t) = 2 \Omega_\LCp \Omega_p(t) + \Omega_\LCm\big(1-\Omega_p^2(t)\big) \;,
 \qquad \text{where}
 	\qquad
 	\Omega_\LCp = \begin{cases}
	0 & t < 0 \\
	\frac{(p-a_\infty) p +m^2}{p_0} & t >0
	\end{cases} \;,
	\quad
	\Omega_\LCm = \begin{cases}
	0 & t < 0 \\
	 -\frac{m a_\infty}{p_0} & t>0
	\end{cases} \;.
\ee
Clearly $\Omega_p(t)\equiv 0\, \forall\, t <0$. While the potential and $\Omega_\LCpm$ jump at $t=0$, the covariance $\Omega_p(t)$ must remain continuous for the Schr\"odinger equation to be obeyed; hence we just need to solve (\ref{delta-schro}) for $t>0$ with the initial condition $\Omega_p(0) = 0$. This is trivial because the $\Omega_\LCpm$ are constants at $t>0$. The solution is
\be
	\Omega_p(t) =  \frac{\Omega_\LCm }{i \pi_0 \cot (\pi_0 t)-\Omega_\LCp} \;,
\ee
which is time-dependent, but there is no reason it should not be. (The covariance must carry, for example, the free energy phases $\sim\exp(-2i\pi_0 t)$ of any created pairs.) To proceed we need the pure gauge vacuum $\ket{\mathbb{0}}$, for which the covariance is $\Omega_p\to \cup_p$ as given in the text. For completeness we write down the fully normalised state:
\be
	\ket{\mathbb{0}} = \exp\bigg[\int\!\ud p  \cup_p b^\dagger_p d^\dagger_{-p} -\frac{V}{2}\log(1+\cup_p^2)\bigg]\ket{0} \;.
\ee
From here we calculate the number of produced pairs from 
\be
	\mathcal{N}(p|a_\infty\theta) = V^{-1} \bra{0;t}B^\dagger_{p+a_\infty} B_{p+a_\infty} \ket{0;t}\bigg|_{t>0} \;,
\ee
finding that at any $t>0$
\be\begin{split}\label{O-minus-U}
	\mathcal{N}(p|a_\infty\theta)
&= \frac{|\Omega_{q}(t)-\cup_q|^2}{(1+\cup_q^2)(1+|\Omega_q(t)|^2)}\bigg|_{q = p+a_\infty}
= \frac{\cup^2_{p+a_\infty}}{1+\cup^2_{p+a_\infty}} \;,
\end{split}\ee
which is time-independent, as it should be. Note that (\ref{O-minus-U}) says that the number of created pairs is supported on the deviation of $\Omega_p$, the solution of the Schr\"odinger equation, from the pure gauge (empty) vacuum, which is sensible. $\mathcal{N}$ is easily evaluated explicitly:
\be\label{delta-dist}
	\mathcal{N}(p|a_\infty\theta) 
	=\frac{1}{2} - \frac{\Omega_+}{2\pi_0} 
	 =\frac{1}{2}-\frac{p a_\infty+p_0^2}{2 p_0 \sqrt{(p+a_\infty)^2+m^2}}
	 \;,
\ee
and is plotted in Fig.~\ref{FIG:DELTA}. This resolves the issues in~\cite{Kim:2011jw}; the unphysical time-dependence encountered there was due only to incorrectly defining physical states at $t>0$ in terms of the Fock vacuum $\ket{0}$, instead of the physical vacuum~$\ket{\mathbb{0}}$. A direct calculation of $\bracket{\mathbb{0}}{0}$ shows that the vacuum persistence probability is also constant at $t> 0$:
\be\label{delta-P-0}
	\mathbb{P}_\text{persist} = \big|\bracket{\mathbb{0}}{0;t}\big|^2\bigg|_{t> 0} = \exp \bigg[- V \int\!\ud p\, \log (1+ \cup_p^2) \bigg] \;.
\ee

\begin{figure}[t!]
	\includegraphics[width=0.5\textwidth]{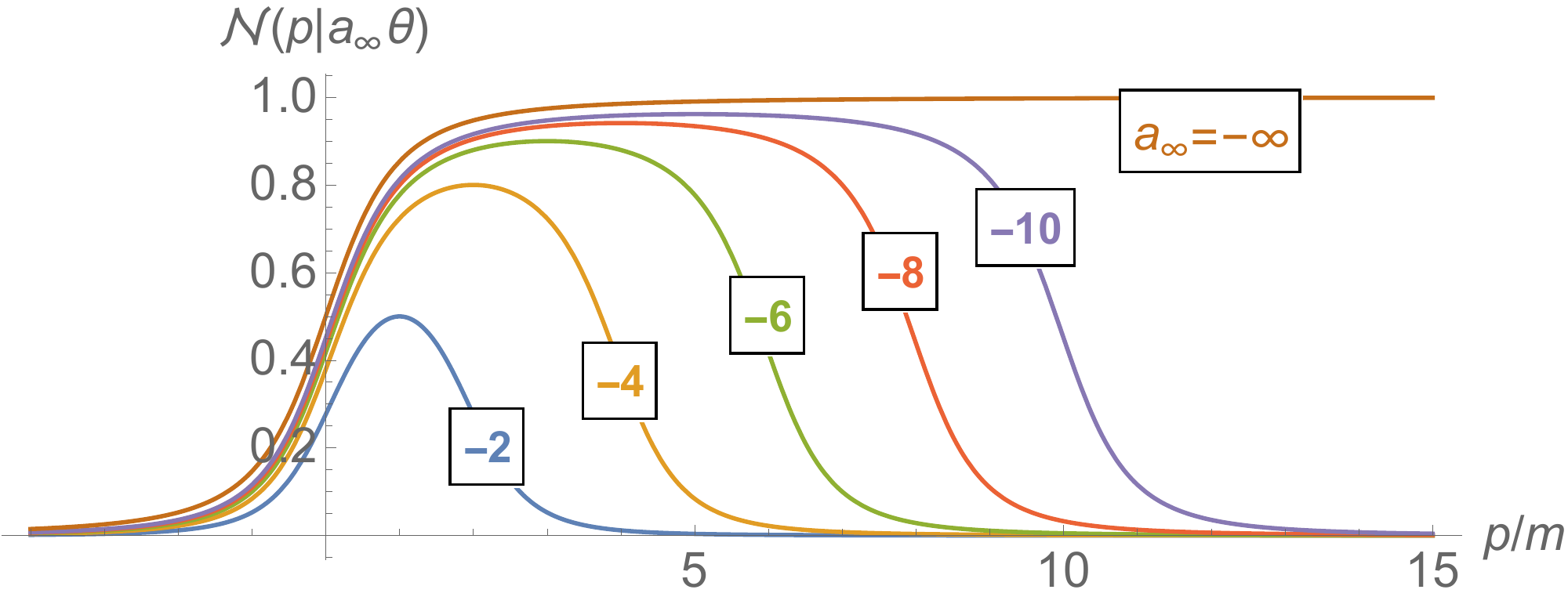}
	\caption{\label{FIG:DELTA} Momentum distribution $\mathcal{N}(p|a_\infty\theta)$ of produced pairs with physical momentum $p$, (\ref{delta-dist}),  created in a delta pulse of various strengths $a_\infty/m = -2,-4\ldots$ as labelled. The distribution becomes broader as $|a_\infty|$ increases, and its peak is at $p=-a_\infty/2$.
}
\end{figure}

\section{Annihlation}\label{AppD}
An applied electric field can produce pairs via the Schwinger effect. The inverse process is that a pair initially present annihilates when an electric field is applied. As this process is perhaps unfamiliar we give some example calculations here. We begin with the initial state $\ket{\text{pair}} = V^{-1}b^\dagger_{p}d^\dagger_{-p}\ket{0}$ at $t=-\infty $describing a pair in the vacuum. Solving the Schr\"odinger equation in a background electric field yields the time-evolved state $\ket{\text{pair};t}$, 
\be\label{ann-state}
	\ket{\text{pair};t}:= \big(\gamma(t) + V^{-1}\Lambda(t) b^\dagger_{p}d^\dagger_{-p}\big)\ket{0;t} \;,
\ee
in which the coefficient functions obey the equations
\be\label{gamma-Lambda}
\begin{split}
	i \dot\Lambda &= 2(\Omega_\LCp-\Omega_\LCm \Omega_p)\Lambda \;, 
	\qquad
	i\dot\gamma = \Omega_\LCm \Lambda  \;, 
\end{split}
\ee
along with the initial conditions $\gamma(t) \to 0$ and $\Lambda(t) \sim e^{-2ip_0 t}$ as $t\to-\infty$. The amplitude for pair annihilation is given by taking the overlap of $\ket{\text{pair};t}$ with the pure gauge vacuum at $t=\infty$. We find
\be\label{ann-overlap}
\bracket{\mathbb{0}}{\text{pair};t} = \bigg(\gamma + \frac{\cup_p \Lambda}{1+ \cup_p \Omega_p}\bigg)\bracket{\mathbb{0}}{0;t} \;.
\ee
This is nonzero in general. To illustrate, we calculate the annihilation probability for the delta-function field in Appendix B. In this case  (\ref{gamma-Lambda}) is easily solved to find
\be
	|\bracket{\mathbb 0}{\text{pair};t}\big|^2 = \cup_p^2 |\bracket{\mathbb 0}{0;t}\big|^2  \qquad \forall\,\, t > 0 \;.
\ee
For completeness we give also an example where the annihilation probability vanishes. We recall the solitonic pair of field and momentum considered in the text, that is $a(t) := (1/ \lambda) \text{sech}\, t/\lambda$ and $p=0$ meaning, here, a pair initially at rest. The covariance $\Omega_{p=0}$ describing the evolution of the vacuum, and the pure gauge vacuum covariance $U_{p=0}$ are
\be
	\Omega_{p=0}(t) = \frac{1}{2\lambda \cosh t/\lambda + i \sinh t/\lambda} \;, \qquad \text{and}\quad \cup_{p=0}(t) = \frac{\sqrt{a(t)^2+1}-1}{a(t)} \;.
\ee
Using these, the solution to (\ref{gamma-Lambda}) is
\be
	\Lambda(t) = \frac{(2 \lambda-i)^2 e^{-2 i t}}{(2 \lambda+i \tanh (\frac{t}{\lambda}))^2} \;, \quad
	\gamma(t) = \frac{(i-2 \lambda) e^{-2 i t}}{(2 \lambda+i)}\Omega(t) \;.
\ee
As $t\to\infty$, we see that $\Omega_{p=0}$, $\cup_{p=0}$,  $\gamma$ and therefore the amplitude (\ref{ann-overlap}) for pair annihilation all go to zero; as such the pair annihilation probability for the solitonic pair is, like the pair creation probability, zero. (\ref{ann-overlap}) suggests that this will be a generic property of solitonic cases. It would be interesting to compare creation and annihilation probabilities in other fields, see e.g.~\cite{Breev:2021lpn} and references therein for analytically tractable cases.

%


%

\end{document}